# Challenger Deep internal wave turbulence events

by Hans van Haren

Royal Netherlands Institute for Sea Research (NIOZ) and Utrecht University, P.O. Box 59, 1790 AB Den Burg, the Netherlands.
hans.van.haren@nioz.nl

ABSTRACT

Marine life has been detected in the ocean's trenches at great depths down to nearly 11 km. Such life is subject to particular environmental conditions of large static pressure exceeding 1000 atmosphere. While current flows are expected to be slow, waters cannot be stagnant with limited exchange of fresh nutrients needed to support life. For sufficient nutrient supply, the physics process of turbulent exchange is required. However, the environmental conditions hamper research in such waters. To study potential turbulent water motions, a string equipped with specially designed high-resolution temperature sensors was moored near the deepest point on Earth in the Challenger Deep, Mariana Trench for nearly three years. A preliminary analysis of a six-day period when the mooring was still demonstrates hundreds of meters slanted convection due to internal waves breaking from above. The associated turbulence dissipation rate with peak values hundred times above the background value is considered sufficient to maintain deep-trench life. Turbulence associates with one-ten thousandth of a degree temperature anomalies of about one hour duration.

*Keywords*: Challenger Deep; long-term mooring; high-resolution temperature measurements; convective turbulent spurs; internal wave driven turbulence

## 1. Introduction

Like in the atmosphere, where breathing would be impossible without turbulent motions, marine life larger than microbial requires turbulent rather than laminar flows for sufficient supply of nutrients and energy, also in the oceans 'hadal' zone of deep trenches (Jamieson, 2015; Gallo et al., 2015; Nunoura et al., 2015). While marine species have been observed at such great depths, establishment and quantification of turbulence processes has been very limited.

Mainly due to the logistical problems imposed by the large hydrostatic pressure which normal oceanographic equipment does not withstand, little is known about the physical oceanography of deep trenches and nothing about the physics that govern the turbulent processes. Turbulence overturn shapes calculated from limited shipborne Conductivity Temperature Depth CTD data from the Puerto Rico Trench averaged over a suitable depth range of 600 m suggest dominant shear-convective turbulence (van Haren, 2015a). Such shear-convective turbulent mixing process is quite different in magnitude from the mainly shear-driven turbulence found in deep passages through ridges and between islands (e.g., Polzin et al. 1996; Lukas et al., 2001; Alford et al., 2011). However, in both cases turbulence is intermittent and has overturn sizes reaching 200 m.

The only moored and hourly sampled measurements so far near the deepest point on Earth, the bottom of the Challenger Deep--Mariana Trench, showed typical current amplitudes of 0.02 m s$^{-1}$ with a dominant semidiurnal tidal periodicity (Taira et al., 2014). Although these authors did not show internal wave band spectra they mentioned sub-peaks at diurnal and inertial frequencies. These data already suggested that trench-waters are not stagnant. Turbulence could not be calculated from these moored observations. In deep lake Baikal, observations (Ravens et al., 2000) from shipborne microstructure profiler indicated very weakly density stratified waters with mean buoyancy frequency $N = 1.4 \times 10^{-4}$ s$^{-1}$, a value which is found around z = -7000 m in the Challenger Deep (van Haren et al., 2017), mean dissipation rates $\varepsilon = O(10^{-10})$ m$^2$ s$^{-3}$ and mean vertical turbulent diffusivities $K_z = 1\text{-}9 \times 10^{-3}$ m$^2$ s$^{-1}$.

In this paper, we report on a six-day detail of high-resolution temperature data from sensors that were moored in the Challenger Deep. The data show internal wave initiated turbulent convection with largest values occurring in a slanted spur-like fashion rather than shear-induced overturning. Near the trench-floor, all dynamics is contained in 0.0001°C temperature changes: A considerable technical challenge to resolve.

## 2. Methods

For a study on internal wave turbulence in hadal zones of the ocean 295 high-resolution temperature T-sensors were custom-made rated to 1400 Bar. Between November 2016 and November 2019 the stand-alone sensors were moored at 11° 19.59′ N, 142° 11.25′ E, about 10,910 m water depth in the Challenger Deep, Mariana Trench, close to the deepest point on Earth (van Haren et al. 2017). The 7 km long mooring consisted of a single pack of floatation providing 2.9 kN net buoyancy at its top 4 km below the ocean surface, more than 6 km of slightly buoyant Dyneema rope as a strength member, and two acoustic releases at 6 m above the anchor-weight. Two sections of instrumentation were in the mooring line: One around 2 km below the top consisting of two current meters and a 200 m long array of standard NIOZ4 T-sensors, and one between 595 and 7 m above the trench floor consisting of the 1400 Bar rated NIOZ4 T-sensors. The single-point Nortek AquaDopp acoustic current meters sampled at a rate of once per 600 s.

All high-resolution NIOZ4 T-sensors (van Haren 2018) were attached at 2 m intervals to 0.0063 m diameter nylon-coated steel cables. The sensors sampled at a rate of once per 2 s and their clocks were synchronized to within 0.02 s via induction every 4 h. The sensors' precision is less than 0.0005°C; the noise level less than 0.0001 °C. Electronic (noise, battery) problems caused failure of about 20% of the sensors after one year and their data were linearly interpolated between neighboring sensors. The extremely weak stratification in the deep $z < -8000$ m waters pushes the sensors to their limits of capability. Post-processing requires a second order pressure correction, besides standard noise and drift correction. The second order pressure correction covers the effects of large-scale surface waves, like tides, and mooring motion of which a vertical variation of 1 m against the local adiabatic lapse rate of $-10^{-4}$ °C m$^{-1}$ shows as 200 m tall artificial internal waves that appear uniform without phase variation over the sensors' vertical range.

For reference, three shipborne SeaBird 911 Conductivity Temperature Depth CTD-profiles were obtained at about 1 km from the mooring site. One profile came to 50 m from the trench-floor (van Haren et al., 2017). The CTD-data are used to establish a temperature-density relationship for quantification of turbulence parameter values from the moored T-sensor data, and for a pressure and drift-correction to the background stratification of the T-sensor data. All analyses were performed after transferring temperature into dynamically correct Conservative (~potential) Temperature ($\Theta$) using the Gibbs-Sea-Water-software described in (IOC, SCOR, IAPSO, 2010).

The moored T-sensor data are used to calculate turbulence dissipation rate $\varepsilon_T = c_1^2 d^2 N^3$ and vertical eddy diffusivity $K_{zT} = m_1 c_1^2 d^2 N$ using the method of reordering potentially unstable vertical density profiles in statically stable ones, as proposed by Thorpe (1977). Here, d denotes the displacements between unordered (measured) and reordered profiles. N denotes the buoyancy frequency computed from the reordered profiles. Rms-values of displacements are not determined over individual overturns, as in Dillon (1982), but over 200 and 588 m vertical intervals that exceed the largest overturn intervals. We use standard constant values of $c_1 = 0.8$ for the Ozmidov/overturn scale factor (Dillon, 1982) and $m_1 = 0.2$ for the mixing efficiency (Osborn, 1980; Oakey, 1982; Ravens et al., 2000), which were mean values in an at least one order of magnitude wide distribution of different turbulence type values established from microstructure profiler data. This is the most commonly used parameterization for oceanographic data after sufficient averaging to achieve statistical stationarity, see further discussions in van Haren et al. (2017) and Gregg et al. (2018). Six-day mean values contain information from about 250,000 profiles and a sufficient mix of samples from shear- and buoyancy-driven turbulence to warrant the above constant values $m_1$ and $c_1$ in the high Reynolds number environment of the deep ocean (Mater et al., 2015; Portwood et al., 2019).

## 3. Observations

The 7-km long mooring line was extending above the surrounding ocean floor at about 5500 m depth and subject to currents that may have been different from those in the trench. Occasionally, this caused mooring motions of a few tens of meters vertically. Because the deep waters in the trench are almost homogeneous with $d\Theta/dz = 5\times10^{-7}$ °C m$^{-1}$ being about 200 times smaller in value and opposite in sign compared to the local adiabatic lapse rate, mooring motions created artificial large internal wave motions. As such artificial internal wave motions are spatially uniform over the vertical, their existence in the data has no effect on turbulent overturning calculations. Nevertheless, to minimize artificial effects and corrections during the post-processing, a six-day section of data is analysed here when the mooring motions were negligible with <0.1 m vertical deflections.

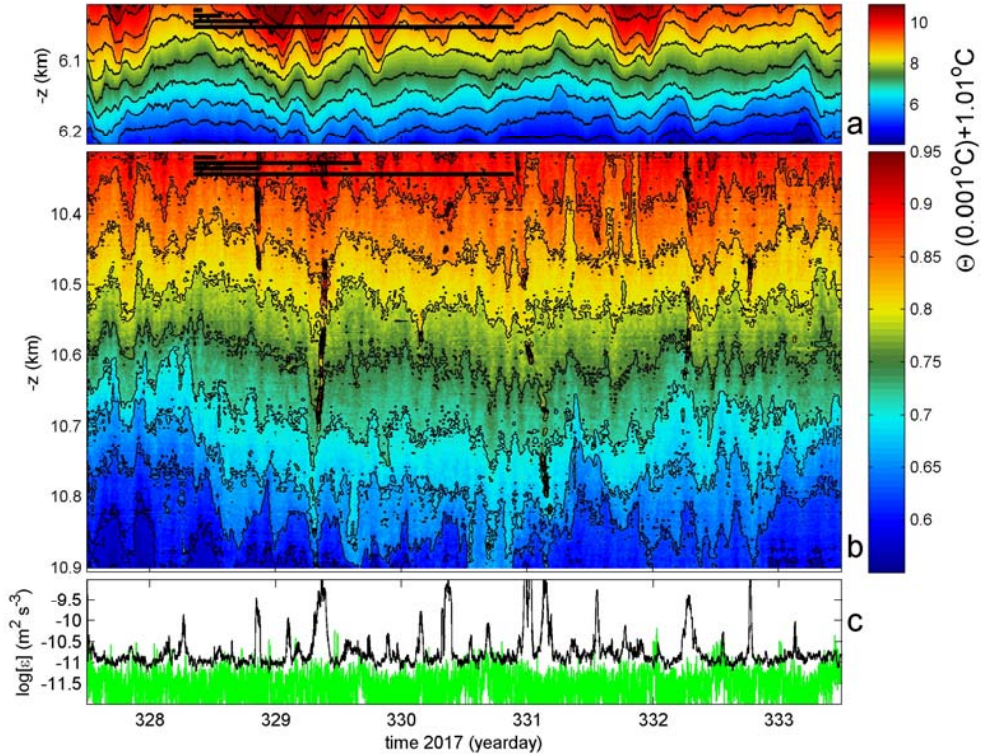

**Figure 1**. Six-day records of high-resolution temperature T-data between 15 and 21 November 2017. (a) Time-depth series for the upper set of T-sensors around z = -6.1 km, low-pass filtered 'lpf' for noise reduction with cut-off at about 150 cpd, short for cycles per day. The black contours are drawn every 0.0007°C. The horizontal bars indicate time-scales of (top-to-bottom): Minimum buoyancy period, mean buoyancy period, semidiurnal period, inertial period. (b) Time-depth series for the deep set of T-sensors around z = -10.6 km, lpf cut-off at 25 cpd. The vertical y-axis range is three times larger than in a., with 100 m tick-marks at identical scale-distances as in a. The trench-floor is at the horizontal x-axis. The colour-range is different than in a. and black contours are drawn every 0.00004°C. (c) Time series of vertically averaged dissipation rates for data in a. (green) and b. (black).

*3.1. Six-day overview*

The high-resolution temperature data reveal omnipresent internal wave activity in the Challenger Deep (Fig. 1). Around z = -6.1 km, isotherm amplitudes are several tens of meters (Fig. 1a). A dominant periodicity does not occur suggesting strong intermittency as is typical for open-ocean internal waves (LeBlond and Mysak, 1978). Every 1.5 to 2.5 days, intensification of amplitudes seems to occur. The 2.5 day periodicity is associated with the local planetary inertial frequency f = 0.39 cpd, short for cycles per day, the lowest internal wave frequency at which internal gravity waves can freely propagate. Turbulent overturning is not directly visible in this graph, although the variable distancing between isotherms suggests internal wave straining and small-scale nonlinear overturning activity: Turbulence is weak but non-negligible (Fig. 1c). Turbulence mean values over 6 days and the 198 m vertical range are, $[<\varepsilon>] = 3.7\pm2\times10^{-12}$ m$^2$ s$^{-3}$, $[<K_z>] = 1\pm0.4\times10^{-5}$ m$^2$ s$^{-1}$, while $<[N]> = 3.3\pm0.5\times10^{-4}$ s$^{-1}$ ($\approx$ 4.5 cpd).

Deeper down around z = -10.6 km near the trench-floor, near-vertical spurs of turbulent overturning anomalies are directly visible from closed contours, e.g. around days 329.3 and 331.1, in the six-day plot (Fig. 1b). In addition, oscillatory motions of isotherms are observed, e.g. around day 330.7 a group of 2 to 3 waves with 2.5 h periodicity. These oscillations cannot be associated with freely propagating internal waves because of the too weak vertical density stratification allowing freely propagating internal waves with periods of larger than one day only. The observed <1 day oscillatory motions have non-uniform amplitudes over the 588 m range of observations. They either represent trapped internal waves, or turbulent overturning. Note that the entire temperature range in this plot is only 0.0004°C, and anomalies are typically 0.0001°C. The anomalies cause the spikes in the vertically averaged turbulence dissipation rate record, reaching values up to $[\varepsilon] = 10^{-9}$ m$^2$s$^{-3}$ for brief periods of about 2 h (Fig. 1c). For these deep data, turbulence values averaged over 6 days and the 588 m vertical range are, $[<\varepsilon>] = 7\pm4\times10^{-11}$ m$^2$ s$^{-3}$, $[<K_z>] = 2.2\pm1\times10^{-3}$ m$^2$ s$^{-1}$, while $<[N]> = 5.9\pm0.8\times10^{-5}$ s$^{-1}$ ($\approx$ 0.8 cpd). It is noted that due to the very weak stratification and T-sensor noise, the threshold for determining $\varepsilon$ is about $6\times10^{-12}$ m$^2$ s$^{-3}$ in this depth range.

The larger turbulence values deeper in the trench associate with the considerably weaker stratification than found around z = -6.1 km and with continued internal wave activity from above facilitating the deep turbulent overturning. While overall correlation between the data around z = -6.1 and -10.6 km is non-significant, visual comparison of Fig. 1a and 1b demonstrates some correspondence between the larger scale periodic motions. Downdraught in isotherms of both data sets is seen around days 329.3 and 331.8.

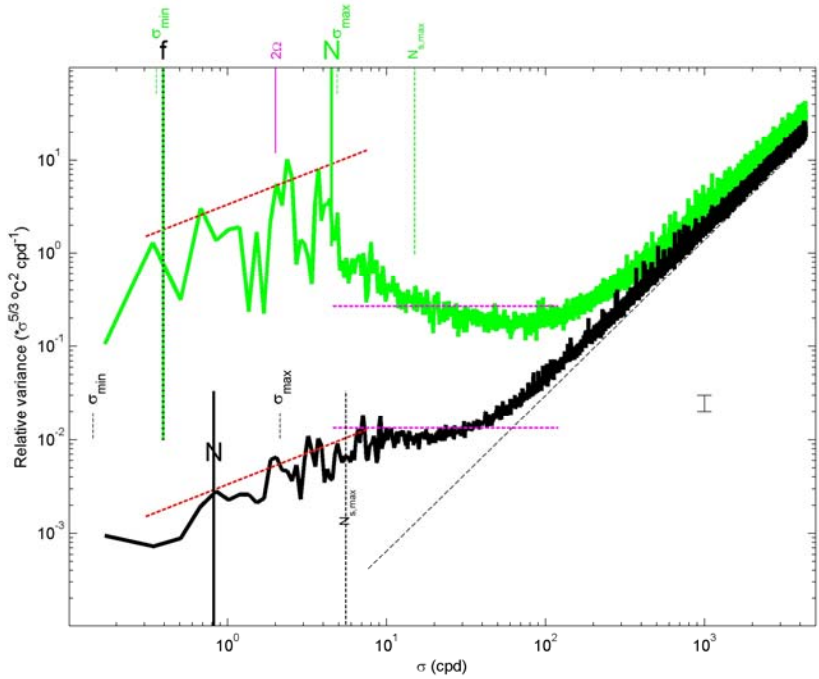

**Figure 2**. Frequency ($\sigma$) plot of data in Fig. 1. Heavily smoothed, about 400 degrees of freedom, vertically averaged temperature variance spectra. The spectra are scaled with the power law $\sigma^{-5/3}$, which reflects the turbulence inertial sub-range. In green, data around z = -6.1 km, in black around z = -10.6 km. Inertial frequency f, semidiurnal Earth rotation $2\Omega$, and the mean N and maximum $N_{max}$ buoyancy frequencies are indicated. Extensions $\sigma_{min} < f$ and $\sigma_{max} > N$ are also indicated. They reflect the internal wave band under weakly stratified conditions accounting for the horizontal Coriolis parameter (see text). The purple dashed line has slope 0 (log-log plot) and represents dominant shear-induced turbulence. The red dashed line has slope +2/3 (log-log plot) and can indicate dominant convective turbulence. The black-dashed line slopes at +5/3 and indicates instrumental white noise.

The weaker temperature stratification in the deep is reflected in more than two orders of magnitude smaller temperature variance and a smaller internal wave band compared to data from around z = -6.1 km (Fig. 2). The log-log spectral plot is scaled with the frequency ($\sigma$) slope of $\sigma^{-5/3}$, which reflects the slope of the turbulent inertial subrange and a mean dominance of shear-driven turbulence or a passive scalar (Tennekes and Lumley, 1972; Warhaft, 2000). Around z = -6.1 km, still great ocean depths of the hadal zone, the inertial subrange is observed for $\sigma > N_{max}$, the maximum 2-m-scale buoyancy frequency in thin layers over the six-day period. Instrumental white noise levels are reached around 150 cpd. Within the internal wave band f ≤ $\sigma \leq N$, the spectrum is weakly increasing and near-horizontal. Upper open-ocean internal wave spectra slope like $\sigma^{-1}$ (van Haren and Gostiaux, 2009), i.e. +2/3 in the plot of Fig. 2. That slope is here mainly observed in the data around $z = -10.6$ km for the super-buoyancy range $1.5 < \sigma < 10$ cpd, where local $N = 0.8$ cpd and $N_{max} = 5.5$ cpd. This frequency band is still partially super-buoyancy when we account for: 1) the horizontal Coriolis force leading to an extended inertia-gravity wave band [$\sigma_{min}<f$, $\sigma_{max}>N$] (LeBlond and Mysak, 1978; Gerkema et al., 2008), and 2) the small-2-m-scale internal wave motions $\sigma < N_{max}$. The spectral slope in this partially super-buoyancy range significantly departs from the slope of the inertial subrange and may also indicate a dominant buoyancy-driven convective turbulence or active scalar (Cimatoribus and van Haren, 2014). For $\sigma >\approx 10$ cpd, the inertial subrange slope is found over a relatively small frequency band, with the notion that instrumental white noise levels are reached around 25 cpd.

*3.2. Detail internal wave turbulence*

A 14 h magnification plot (Fig. 3) demonstrates the internal waves and turbulent overturning events near the trench floor. Isotherm excursions are up to 100 m crest to trough. Shear-driven isotherm deformation leading to overturning is visible, e.g. around $z = -10.45$ km and day 331.0. Also, non-wave like temperature anomalies of relatively colder water surrounded by relatively warmer water are observed in sloping layers slanted to the vertical, and also near the trench floor, extending in a group over 100 m in the vertical, e.g., around day 331.35. However, the spectral information of Fig. 2 suggests a dominant convective overturning. A large, presumably convective, slanted spur of overturning warm anomalies is observed between day 330.9 $z = -10.3$ km and day 331.2 $z = -10.8$ km. It is split in smaller anomalies. The overturn anomalies slope at various angles. The small ones have vertical extents of 10 to 20 m with a duration of about 1 h, e.g. around day 331.15 and $z = -10.67$ km, and 40 m lasting half an hour around day 331.15 and $z = -10.75$ km. For an estimated ambient current speed of 0.02 m s$^{-1}$, a 1 h duration translates to 75 m horizontal distance, which yields slanting slope aspect ratios varying between 0.2 and 1.2.

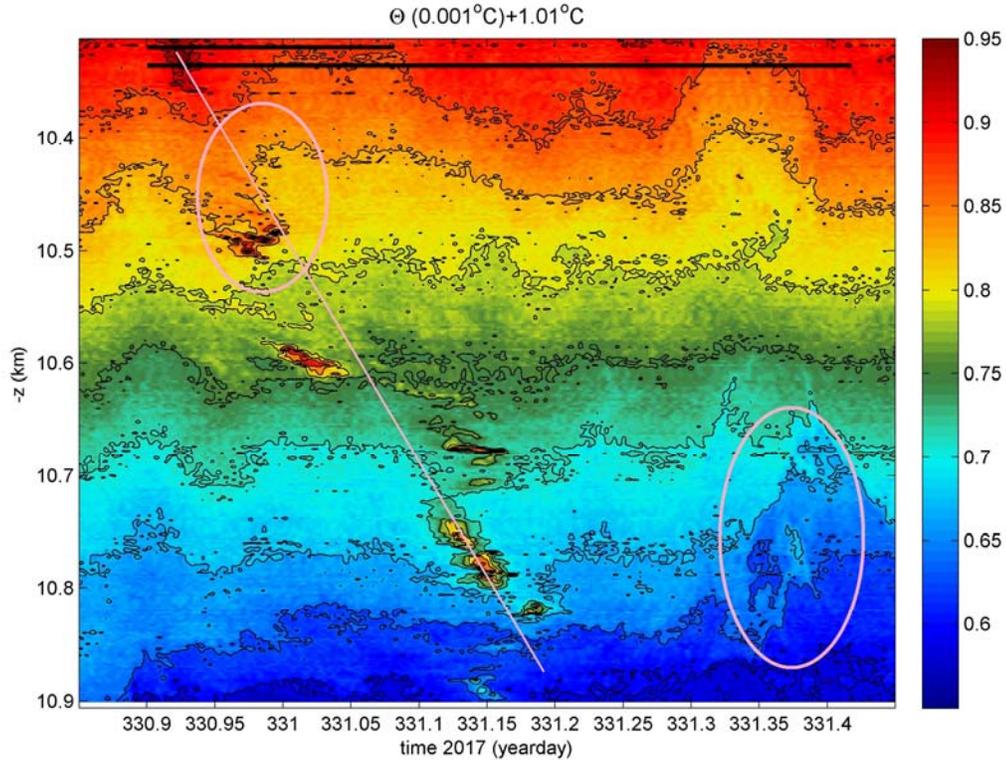

**Figure 3**. 14 h detail of Fig. 1b, around the passage of a relatively strong turbulence peak above the trench floor. Horizontal bars only indicate the minimum buoyancy and the semidiurnal periods.

## 4. Discussion and concluding remarks

The computed turbulence values compare with open-ocean values (e.g., Gregg, 1989; Polzin et al., 1997). Near the trench-floor turbulence values are found one to two orders of magnitude larger compared with values in the more stratified waters near the trench-top. The larger values in the weaker stratified deep trench waters are several orders of magnitude above molecular diffusion levels. The larger turbulence values are all attributable to about 1 h duration spurs of 0.0001°C warm anomalies with respect to local environmental temperatures. Smaller spurs of 10 to 40 m vertical extent organize in 500 m tall ones that are slanted to the vertical. The small spurs are also slanted to the vertical at a wide variety of angles.

Such slanted convection is known to occur in an environment with a background sheared current-flow, or, mainly in very weakly stratified waters, due to the effects of the horizontal component of the Coriolis force (Straneo et al., 2002; Sheremet, 2004). While the former can have a wide range of slopes, the latter have a preferential direction of the Earth rotational vector. At the latitude of the mooring, the Earth vector has a local slope of $\tan(11.3°) = 0.2$ to the horizontal. This slope is comparable to some of the observed convection spur slopes. The temperature anomaly suggests a source 200 m higher-up, if coming from above against the background stratification, or a source sideways from the trench-walls. Both shear and Earth rotational processes leading to slanted convection still need a process that initiates the convection. Such initiating of convection bursts is not well established for a stratified environment. Recent observations suggested internal wave accelerations into weakly stratified waters from above to generate such convection (van Haren, 2015b). The present observations may lead to the same conclusion, albeit convection here occurs in shorter bursts that eventually follow slanted pathways. This is concluded from the spectra around $z = -10.6$ km pointing at dominant convective motions in the super-buoyancy range, with oscillatory isotherm motions that are coupled to free internal wave propagation higher up around $z = -6.1$ km. These observations demand future refinements of internal wave-turbulence modelling.

The turbulence values are expected to be sufficient for replenishment of nutrients and suspended materials, depending on available sources. Trenches collect elevated deposition of organic matter relative to the surrounding ocean floor, which results in twice larger biological oxygen consumption (Glud et al., 2013). Sources and their replenishment are thus expected to be abundant for turbulent redistribution, also near the trench floor.

*Acknowledgements.* I thank the masters and crews of the R/V Sonne and R/V Sally Ride for the pleasant cooperation during the operations at sea. I am grateful for all the support by Martin Laan and Yvo Witte.

**Competing interests** The author declares no competing interests.

**Figure 1**. Six-day records of high-resolution temperature T-data between 15 and 21 November 2017. (a) Time-depth series for the upper set of T-sensors around z = -6.1 km, low-pass filtered 'lpf' for noise reduction with cut-off at about 150 cpd, short for cycles per day. The black contours are drawn every 0.0007°C. The horizontal bars indicate time-scales of (top-to-bottom): Minimum buoyancy period, mean buoyancy period, semidiurnal period, inertial period. (b) Time-depth series for the deep set of T-sensors around z = -10.6 km, lpf cut-off at 25 cpd. The vertical y-axis range is three times larger than in a., with 100 m tick-marks at identical scale-distances as in a. The trench-floor is at the horizontal x-axis. The colour-range is different than in a. and black contours are drawn every 0.00004°C. (c) Time series of vertically averaged dissipation rates for data in a. (green) and b. (black).

**Figure 2**. Frequency ($\sigma$) plot of data in Fig. 1. Heavily smoothed, about 400 degrees of freedom, vertically averaged temperature variance spectra. The spectra are scaled with the power law $\sigma^{-5/3}$, which reflects the turbulence inertial sub-range. In green, data around z = -6.1 km, in black around z = -10.6 km. Inertial frequency f, semidiurnal Earth rotation $2\Omega$, and the mean N and maximum $N_{max}$ buoyancy frequencies are indicated. Extensions $\sigma_{min}$ < f and $\sigma_{max}$ > N are also indicated. They reflect the internal wave band under weakly stratified conditions accounting for the horizontal Coriolis parameter (see text). The purple dashed line has slope 0 (log-log plot) and represents dominant shear-induced turbulence. The red dashed line has slope +2/3 (log-log plot) and can indicate dominant convective turbulence. The black-dashed line slopes at +5/3 and indicates instrumental white noise.

**Figure 3**. 14 h detail of Fig. 1b, around the passage of a relatively strong turbulence peak above the trench floor. Horizontal bars only indicate the minimum buoyancy and the semidiurnal periods.

15